\documentclass[onecolumn,showpacs,nobibnotes,nofootinbib, greek, hebrew,12pt ]{revtex4}
\usepackage[ddmmyy,hhmmss]{datetime}
\usepackage{setspace}
\usepackage{amsmath,amssymb}
\usepackage{graphicx}
\usepackage{subfigure}
\usepackage{graphicx,amssymb}

\newcommand{\be}{\begin{equation}}
\newcommand{\ee}{\end{equation}}
\newcommand{\bea}{\begin{eqnarray}}
\newcommand{\eea}{\end{eqnarray}}

\def\beq{\begin{equation}}
\def\eeq{\end{equation}}

\def\beqn{\begin{eqnarray}}
\def\eeqn{\end{eqnarray}}

\begin{document}

\title{  Hermitian separability and transition from  singlet  to  adjoint BFKL equations in $\mathcal{N}=4$ super Yang-Mills Theory}
\author{ S.~Bondarenko$^{(1) }$ and A.~Prygarin$^{(1),(2)}$ }
\affiliation{ $^{(1) }$Physics Department, Ariel University, Ariel 40700, Israel \\
$^{ (2)}$II. Institute of Theoretical Physics, Hamburg University, Germany}
\pacs{12.60.Jv, 12.38.-t, 11.38.Bx}

\singlespacing
\small
\begin{abstract}
We revisit the next-to-leading order~(NLO) correction to the eigenvalue of the BFKL equation in the adjoint representation and  investigate its properties  in analogy with the singlet BFKL  in planar $\mathcal{N}=4$ super Yang-Mills Theory~(SYM). We show that the  adjoint NLO  BFKL eigenvalue is needed to be slightly modified in order to have a  property of  hermitian separability present for the singlet BFKL. After this modification the  adjoint NLO  BFKL eigenvalue is expressed through holomorphic and antiholomophic parts of the leading order eigenvalue and their derivatives. The proposed choice of the modified NLO expression  is supported by the fact that it is possible to obtain the same result in a relatively straightforward way directly  from the  singlet NLO BFKL eigenvalue replacing alternating   sums by non-alternating ones. This transformation corresponds to changing cylindrical topology of the singlet BFKL to the planar topology of the adjoint BFKL. We believe that the original NLO calculation of Fadin and Lipatov is correct and valid for the computations of the remainder function of the BDS amplitude. However, the notion of the  adjoint BFKL eigenvalue is vaguely  defined   due to removal of the infrared divergences  as well as redistributing NLO corrections between the kernel and impact factors, and  is to be modified  to comply with properties of the singlet BFKL equation. This, at first sight, a purely semantic difference may become important in resolving the issue of the non-vanishing adjoint NNLO eigenvalue in the limit of zero anomalous dimension $\nu$ and conformal spin $n$, which contradicts the bootstrap condition of the BFKL equation.
\end{abstract}
 \maketitle
\normalsize

\section{Introduction}
In the context of the Regge theory the scattering amplitude at large energy $\sqrt{s}$ and fixed momentum transfer $\sqrt{-t}$ can be written as follows~\cite{Gribov}
\beqn
A^p(s,t) =\xi_p(t) s^{1+\omega_p(t)} \gamma^2(t), \xi_p(t)=e^{-i \pi \omega_p(t)}-p,
\eeqn
where $p=\pm 1$ is the signature of the reggeon with trajectory $\omega_p(t)$ and $\gamma^2(t)$ stands for the product of reggeon-projectile vertices. The leading Regge trajectory with vacuum quantum numbers is associated with Pomeron and has approximately constant behavior of the total cross section for hadron-hadron scattering. According to the Mandelstam argument Pomeron generate cut singularities in the $\omega$-plane~\cite{Mandelstam}.
In the leading logarithmic order of QCD, the Regge factorization of the scattering amplitude allows to construct the Balitski-Fadin-Kuraev-Lipatov~(BFKL) equation from the leading order multiparticle production amplitudes using analyticity, unitarity, renormalizability and crossing symmetry~\cite{BFKL}.

The integral kernel of the BFKL equation possesses remarkable properties of holomorphic separability~\cite{int1} and invariance under M\"{o}bius transformation~\cite{moeb}.  Its generalized form to a composite state of  a number of reggeized  gluons, known as Bartels-Kwiecinski-Praszalowicz~(BKP) equation~\cite{BKP}, in the t'Hooft limit leads to an integrable Heisenberg spin chain model~\cite{int, integrQP} possessing duality symmetry~\cite{dual}.

The next-to-leading~(NLO) corrections to the singlet BFKL equation  in QCD were calculated in~\cite{FL} containing non-analytic terms proportional to the Kronecker symbol of the conformal spin of the M\"{o}bius  group, which cancel in $\mathcal{N}=4$ SYM leading to the hermitian separability~\cite{KotikovDGLAP} and the maximal transcendentality property of the singlet NLO BFKL eigenvalue~\cite{KL, trajN4}. The property of the maximal transcendentality is now commonly  believed to be a general property of the $\mathcal{N}=4$ SYM perturbative loop calculations, in particular it was very useful in calculating the anomalous dimension of the twist-two operators~\cite{KLOV, KLOV1}.

The BFKL approach was shown  to be very useful in checking the analytic properties of the Bern-Dixon-Smirnov~(BDS) amplitude~\cite{BDS}   and made it possible to calculate the leading-logarithmic correction of the so called remainder function~\cite{sabio1, sabio2}, which led to a series of papers extending this analysis to subleading corrections, connection between Regge and collinear kinematics as well as beyond MHV helicity configurations of the external particles~\cite{LP1,LP2, BLPreview, 3to3, BLPcollinear, PSVV, BKLP, LPS, korm1, korm2, dixon3loops, Dixon2012yy, Pennington2012zj, CaronHuot2013fea,Hatsuda2014oza, Dixon4loops, BCHS, JamesYorgos, Dixon2014xca,Dixon2014iba, POPE}. The Regge kinematics of the BDS amplitude  at the strong coupling side was investigated in \cite{kot1,spr1,kot2,spr2,spr3}.

The next-to-leading~(NLO) correction to the eigenvalue of the BFKL equation in the adjoint representation was calculated in \cite{FadinAdj}, which allowed to fix some free parameters in the three loop ansatz of the BDS remainder function in general kinematics~\cite{dixon3loops}. The M\"{o}bius invariance of the kernel of the adjoint  NLO BFKL equation was studied in  \cite{moebadj}.


In this article we study the property of the hermitian separability of the next-to-leading order correction to the eigenvalue of the BFKL equation in the color adjoint representation in the planar limit of the infinite number of colors in $\mathcal{N}=4$ SYM.
It was shown by Kotikov and Lipatov~\cite{KotikovDGLAP} that the singlet NLO BFKL eigenvalue does not have holomorphic separability present at the leading order, but it  endows a property of the hermitian separability if written in the form of Bethe-Salpeter equation. In attempt of applying the same analysis to the adjoint NLO BFKL eigenvalue found in \cite{FadinAdj}, we face a problem that can be resolved by slightly modifying the NLO eigenvalue by adding $\psi$-functions to some terms. After this modification the adjoint NLO BFKL eigenvalue is readily brought to the form in which the property of hermitian separability becomes manifest (see (\ref{E1tilde})) and it is a function only of the holomorphic and antiholomorphic part of the leading order eigenvalue and their derivatives.  The suggested modification of the  adjoint NLO BFKL eigenvalue is supported by the fact that it is possible to find a transformation, which takes the singlet NLO BFKL eigenvalue to the adjoint one  reproducing it in the modified form suggested in the present article. This transformation is naturally derived if one notices that the appearance of alternating nested sums in the singlet NLO eigenvalue is related to the cylindrical topology as was observed in QCD calculations of the double logarithms~\footnote{We thank to Lev Lipatov for bringing  our attention to this fact.}. Replacing $(-1)^k$ factor in the sum over $k$ by unity together with the shift of the argument of the corresponding $\psi$-function by $1/2$ reproduces the adjoint NLO BFKL eigenvalue in the modified form compatible with the property of the hermitian separability.
  We believe that the result of \cite{FadinAdj} is not in contradiction with our findings and still is valid for calculations of the BDS remainder function, but the notion of the BFKL eigenvalue in the adjoint representation should be changed in favor of the modified form presented in this article. This is due the fact that in contrast to the well defined singlet BFKL equation, the definition of the BFKL  eigenvalue     in the adjoint representation largely depends on the way one removes the infrared singularities and associated finite terms already preset in the BDS amplitude, not to mention some ambiguity in redistributing the beyond-leading order corrections between the BFKL eigenvalue and the impact factors.

  The present article is organized as follows.
  The first section is devoted to the analysis of the hermitian separability of the adjoint NLO BFKL eigenvalue and argue that in order to comply with this property the adjoint NLO BFKL eigenvalue is to be slightly modified. This modification is not supposed to affect the final result for the remainder function at this level, but might be important for higher loops. In the second section we propose a  direct way of  deriving  the modified adjoint BFKL eigenvalue from the well known singlet NLO BFKL eigenvalue, supporting the results of the first section.
  A brief summary of our arguments is presented in Conclusions.

\section{Hermitian separability}\label{sec:hermit}

 For our analysis of the next-to-leading order~(NLO) BFKL eigenvalue in the adjoint representation~\cite{FadinAdj} we adopt the notation of Dixon, Drummond, Duhr and Pennington~\cite{Dixon4loops}.
The maximally helicity violating~(MHV) amplitude  $A_6^{MHV}$ can be written in the form of the Bern-Dixon-Smirnov aplitude $A_6^{BDS}$~\cite{BDS} corrected by a term named the remainder function, which becomes important already at two loops in the MHV amplitude with six external particles
\beqn
A_6^{MHV}=A_6^{BDS} \times \exp(R_6).
\eeqn
The full amplitude is infrared divergent and all divergencies are captured by the BDS amplitude leaving the remainder function infrared finite. The remainder function is conformal invariant in the space of dual coordinates $p_i=x_i-x_{i+1}$~\cite{DualConformal, FourLoopNeq4, FiveLoopNeq4, AMStrong, Drummond2008vq, Bern2008ap}, which makes it to be a function of the three dual conformal cross ratios for the six particle amplitude,
\beqn
u_1=\frac{x^2_{13} x^2_{46}}{x^2_{14} x^2_{36}}, \;
u_2=\frac{x^2_{24} x^2_{51}}{x^2_{25} x^2_{41}},
\;
u_3=\frac{x^2_{35} x^2_{62}}{x^2_{36} x^2_{52}}.
\eeqn

In the special case of the multi-Regge kinematics the cross ratios take values
\beqn
u_1 \to 1, \; u_2, u_3 \to 0
\eeqn
possessing phases in some kinematical regions~\cite{sabio1, sabio2}, named Mandelstam regions, where  the ratios
\beqn\label{ww}
\frac{u_2}{1-u_1} \equiv \frac{1}{(1+w)(1+w^*)}, \;\; \frac{u_3}{1-u_1} \equiv \frac{w w^*}{(1+w)(1+w^*)}
\eeqn
held fixed. The complex variable $w$  in are introduced in (\ref{ww}) to eliminate square roots of the linear combination of $u^2_i$ and to make the target-projectile symmetry explicit  in the final expression of the remainder function in the Mandelstam region of the $2 \to 4$ scattering amplitude,
\beqn\label{R6}
&&e^{R_6+i \pi \delta}|_{MRK}= \cos \pi \omega_{ab} +i \frac{a}{2} \sum_{n=-\infty}^{\infty} (-1)^n \left(\frac{w}{w^*}\right)^{\frac{n}{2}}
\int_{-\infty}^{\infty} \frac{d \nu }{\nu^2 +\frac{n^2}{4}}|w|^{2 i \nu} \Phi_{Reg}(\nu, n)  \nonumber
 \\
 &&\hspace{4cm}  \times\exp \left[ -\omega^{a}(\nu,n)\left(\ln(1-u_1)+i \pi+\frac{1}{2}  \ln \frac{|w|^2}{|1+w|^4}\right)\right],
\eeqn
where
\beqn\label{omegaab}
\omega_{ab}=\frac{1}{8} \gamma_K(a) \ln |w|^2, \; \; \delta=\frac{1}{8}\gamma_K(a) \ln \frac{|w|^2}{|1+w|^4}
\eeqn
represent contribution of the Regge poles and  some finite pieces of the Mandelstam~(Regge) cut already present in the BDS amplitude.
The factor $\gamma_K$ stands for the cusp anomalous dimension known at any value of the coupling constant.
The contribution of the  Mandelstam cut enters the remainder function through solution of the BFKL equation in the adjoint representation, which eigenvalue reads
\beqn\label{omegaadjoint}
\omega^{adjoint}(\nu,n )=-a \left(E_{\nu,n}+a E^{(1)}_{\nu,n}+\mathcal{O}(a^2)\right).
\eeqn
The leading order eigenvalue  is given by
\beqn\label{Ehol}
E_{\nu,n}=-\frac{1}{2} \frac{|n|}{\nu^2 +\frac{n^2}{4}}+\psi\left(1+i \nu+\frac{|n|}{2}\right)
+\psi\left(1-i \nu+\frac{|n|}{2}\right)-2 \psi(1)=E^{+}_{\nu,n}+E^{-}_{\nu,n}
\eeqn
in terms  of its  holomorphic  and antiholomorphic part ,
\beqn
E^{\pm}_{\nu,n}\equiv  -\frac{1}{2} \frac{1}{ \pm i\nu +\frac{|n|}{2}}+\psi\left(1 \pm i \nu+\frac{|n|}{2}\right)- \psi(1).
\eeqn
Here $\psi(z)=\frac{d}{d z} \ln \Gamma(z)$ is the digamma function with $\psi(1)=-\gamma_E$ being the Euler-Mascheroni constant.
The next-to-leading~(NLO) adjoint BFKL eigenvalue was calculated in \cite{FadinAdj} and can be conveniently written as follows~\cite{Dixon4loops}
\beqn\label{E1Dixon}
E^{(1)}_{\nu,n}= -\frac{1}{4}D^2_{\nu} E_{\nu,n} +\frac{1}{2} V D_{\nu} E_{\nu,n}-\zeta_2 E_{\nu,n}-3 \zeta_3
\eeqn
in terms of
\beqn\label{VV}
V \equiv -\frac{1}{2} \left[\frac{1}{i \nu +\frac{|n|}{2}}-\frac{1}{-i \nu +\frac{|n|}{2}}\right]=\frac{i \nu}{\nu^2 +\frac{n^2}{4}}
\eeqn
and
\beqn
N \equiv \text{sgn}(n)\left[\frac{1}{i \nu +\frac{|n|}{2}}+\frac{1}{-i \nu +\frac{|n|}{2}}\right] =\frac{n}{\nu^2+\frac{n^2}{4}}
\eeqn
as well as derivative  defined by $D_\nu=-i \partial_\nu$.
The NLO corrections to the product of the two impact factors are encoded in the function
\beqn
\Phi_{Reg}(\nu,n )=1+a \Phi^{(1)}_{Reg}(\nu,n )+a^2 \Phi^{(2)}_{Reg}(\nu,n )+\mathcal{O}(a^3),
\eeqn
where the subleading   terms are given by
\beqn
\Phi^{(1)}_{Reg}(\nu,n)= -\frac{1}{2} E^2_{\nu,n}-\frac{3}{8} N^2-\zeta_2
\eeqn
and
\beqn
&& \Phi^{(2)}_{Reg}(\nu,n) =\frac{1}{2} \left[\Phi^{(1)}_{Reg}(\nu,n)\right]^2 -E^{(1)}_{\nu,n}E_{\nu,n}+\frac{1}{8} \left[D_\nu E_{\nu,n}\right]^2+\frac{5}{64} N^2 \left(N^2+4V^2\right) \nonumber
\\
&&\hspace{2cm}-\frac{\zeta_2}{4} \left(2 E^{2}_{\nu,n}+N^2+6 V^2\right)+\frac{17}{4} \zeta_4.
\eeqn

It was shown by Kotikov and Lipatov~\cite{KotikovDGLAP} that the  Schr\"{o}dinger type  BFKL eigenvalue has a more transparent meaning if written in the form of the  Bethe-Salpeter equation at the next-to-leading order as follows
\beqn
1=\frac{a f_1 +a^2 f_2 }{\omega^{adjoint}(\nu,n )} +ag_1
\eeqn
or equivalently
\beqn\label{omegaf1g1}
\omega^{adjoint}(\nu,n )= a f_1 +a^2 (f_2 +g_1 f_1)+\mathcal{O}(a^3).
\eeqn

The functions $f_i$ and $g_1$  should be of  the following form
\beqn
f_i=f_i^{+}+f_i^{-}, \; g_1=g_1^{+}+g_1^{-},
\eeqn
where $f_i^{\pm}$ and $g_1^{\pm}$ have poles in either   upper or lower part of the complex  $\nu$-plane. This is what we call Hermitian separability of the BFKL eigenvalue.

In this picture the NLO BFKL eigenvalue is written using simpler functions with holomorphic and antiholomorphic parts being separated.  We believe that reformulation of calculating   the next-to-leading corrections to the BFKL eigenvalue in terms  of these functions can be useful in context of novel integrability techniques, e.g. Quantum Spectral Curves etc.

Applying  (\ref{omegaf1g1}) directly to the  (\ref{E1Dixon}) one immediately faces a problem of decomposing it in terms of  $f_i^{\pm}$ and $g_1^{\pm}$ with upper and lower poles being separated. This fact makes it difficult to translate all the properties of the singlet BFKL equation to the adjoint BFKL. One way to resolve it is to modify the problematic expression for $V$ in  (\ref{VV}), which causes the difficulty of writing it in the form of (\ref{omegaf1g1}).
We propose to replace $V $  in  (\ref{E1Dixon})  by  the following expression
\beqn\label{Vtilde}
\tilde{V}\equiv-\frac{1}{2}  \frac{1}{i \nu +\frac{|n|}{2}}+\psi\left(1 + i \nu+\frac{|n|}{2}\right)+\frac{1}{2}\frac{1}{-i \nu +\frac{|n|}{2}}-\psi\left(1 - i \nu+\frac{|n|}{2}\right)=E^{+}_{\nu,n}-E^{-}_{\nu,n},
\eeqn
which seems to be very natural if written in terms of $E^{\pm}_{\nu,n}$. This modification does not affect the following limits at the next-to-leading   level
\beqn
\lim_{\nu \to 0} \omega^{adjoint}(\nu,0)=0, \;\;
\lim_{\nu \to \infty} \omega^{adjoint}(\nu,n) \simeq \gamma_K \ln \nu^2.
\eeqn

Substituting $V$ by $\tilde{V} =E^{+}_{\nu,n}-E^{-}_{\nu,n}$ we can write $E^{(1)}_{\nu,n}$ as follows
 \beqn\label{E1tilde}
&& \tilde{E}^{(1)}_{\nu,n}= -\frac{1}{4}D^2_{\nu}(E^{+}_{\nu,n}+E^{-}_{\nu,n}) +\frac{1}{2} (E^{+}_{\nu,n}-E^{-}_{\nu,n}) D_{\nu} (E^{+}_{\nu,n}+E^{-}_{\nu,n})-\zeta_2 (E^{+}_{\nu,n}+E^{-}_{\nu,n})-3 \zeta_3 \nonumber
\\
&&
=-\frac{1}{4}D^2_{\nu}(E^{+}_{\nu,n}+E^{-}_{\nu,n}) +\frac{1}{2}D_{\nu}\left( \left(E^{+}_{\nu,n}\right)^2-\left(E^{-}_{\nu,n}\right)^2\right)-\frac{1}{2} (E^{+}_{\nu,n}+E^{-}_{\nu,n}) D_{\nu} (E^{+}_{\nu,n}-E^{-}_{\nu,n}) \nonumber
\\
&&\hspace{1cm}-\zeta_2 (E^{+}_{\nu,n}+E^{-}_{\nu,n})-3 \zeta_3.
\eeqn
 From (\ref{omegaadjoint}), (\ref{Ehol}) and  (\ref{omegaf1g1}) one readily   spells out the functions
  \beqn
 f_1^{\pm}=-E^{\pm}_{\nu,n}, \;\; g_1^{\pm}= \mp\frac{1}{2} D_{\nu} E^{\pm}_{\nu,n} - \zeta_2
 \eeqn
 and
 \beqn
 f_2^{\pm}= \frac{1}{4}D^2_{\nu} E^{\pm}_{\nu,n} \mp \frac{1}{2} D_{\nu} \left(E^{\pm}_{\nu,n}\right)^2 +\frac{3}{2} \zeta_3.
 \eeqn

 At first sight it is impossible to add by hand arbitrary terms to the BFKL eigenvalue. However, we recall that there is a freedom in   redistributing corrections between the eigenvalue and the impact factors as well some uncertainty due to removing infrared divergences from the infra-red divergent adjoint BFKL equation.   We take advantage of those two facts and argue that is it more appropriate to call the  next-to-leading adjoint BFKL eigenvalue, the one with $V \to \tilde{V}=E^{+}_{\nu,n}-E^{-}_{\nu,n}$ in  (\ref{E1Dixon}). In other words if one wishes to translate the properties of the singlet BFKL equation, like hermitian separability and bootstrap condition, to the adjoint BFKL than a  proper object to consider is the modified expression $\tilde{E}^{(1)}_{\nu,n}$ in (\ref{E1tilde}).
 To our opinion this redefinition is only needed for checking properties of the  BFKL eigenvalue and  does not affect calculations of the remainder function in (\ref{R6}), where one still should use (\ref{E1Dixon}) with (\ref{VV}). Alternatively, due to the arbitrariness in the choice of the energy scale~(see \cite{fadin_scale})  one can use in (\ref{R6}) the redefined  eigenvalue of   (\ref{E1tilde}) accompanied by a proper redefinition of the impact factor and the energy scale, while keeping the result for the amplitude and thus for the remainder function unchanged.

 Our proposal is supported by a connection between the  singlet and adjoint  BFKL eigenvalues   presented in the next section.

\section{From Singlet to  Adjoint BFKL}\label{fromto}
The NLO corrections to the BFKL equation are   known for about two decades. Due to   numerous functional identities it has several equivalent  representations. Here we adopt the one used in ref.~\cite{KotikovDGLAP} because it is closely related to the discussion of the previous section and allows clarity of presentation. The BFKL eigenvalue can be written as
 \beqn\label{omegasinglet}
\omega^{singlet}(n,\nu)= 4 a \left[\chi(n,\gamma)+a \; \delta(n,\gamma)+\mathcal{O}(a^2)\right],
\eeqn
where the leading order term and the next-to-leading correction are given by
\beqn
\chi(n,\gamma)=2 \psi(1) -\psi(M)-\psi(1-\tilde{M})
\eeqn
as well as
\beqn\label{delta}
\delta(n, \gamma)=\psi''(M)+\psi''(1-\tilde{M})+6 \zeta_3-2 \zeta_2 \chi(n,\gamma)-2\Phi(|n|,\gamma)-2\Phi(|n|,1-\gamma).
\eeqn
The LO and NLO terms are functions of  holomorphic and antiholomorphic variables
\beqn
M=\gamma+\frac{|n|}{2}, \; \tilde{M}=\gamma-\frac{|n|}{2}, \;\gamma=\frac{1}{2}- i\nu,
\eeqn
where $\gamma$ and $n$ are the anomalous dimension  and the conformal spin respectively.
The most complicated part of the NLO expression can be written as follows~\cite{KotikovDGLAP}
\beqn\label{PhiGamma}
&&\Phi(|n|,\gamma)+\Phi(|n|,1-\gamma)=\chi(n,\gamma) \left(\beta'(M)+\beta'(1-\tilde{M})\right) \nonumber
\\
&&+
\Phi_2(M)-\beta'(M)\left[\psi(1)-\psi(M)\right]+
\Phi_2(1-\tilde{M})-\beta'(1-\tilde{M})\left[\psi(1)-\psi(1-\tilde{M})\right] \nonumber
\\
&&
= \Phi_2(M)+\beta'(M)\left[\psi(1)-\psi(1-\tilde{M})\right]+
\Phi_2(1-\tilde{M})+\beta'(1-\tilde{M})\left[\psi(1)-\psi(M)\right]
\eeqn
in terms of
\beqn\label{Phi2}
\Phi_2(z)=\sum_{k=0}^{\infty} \frac{\beta'(k+1)+ (-1)^k \psi'(k+1)}{k+z}-\sum_{k=0}^{\infty} \frac{ (-1)^k (\psi(k+1)-\psi(1))}{(k+z)^2}
\eeqn
and
\beqn\label{betaprime}
\beta'(z)=-\sum_{k=0}^{\infty} \frac{(-1)^k}{(k+z)^2}.
\eeqn
Here  we notice      alternating sums in contrast to the  the adjoint BFKL eigenvalue in (\ref{E1Dixon}), which can be written using only constant sign sums. The alternating sums appear due to a non-planar cylindrical topology of the singlet BFKL equation compared to the planar color adjoint BFKL. Both the singlet and the adjoint BFKL equations are integrable corresponding to   closed and open Heisenberg spin chains respectively.
This similarity becomes most obvious at the leading order, where the two eigenvalues are expressed using the same $\psi$-function with an argument shifted by $1/2$. Namely, the singlet BFKL leading order reads
\small
\beqn\label{chi01}
&& \chi(n, \gamma)=2 \psi(1) -\psi(M)-\psi(1-\tilde{M})
\\
&&=-\frac{1}{2} \left(\psi\left(\frac{1}{2}+i\nu +\frac{n}{2}\right)
+
\psi\left(\frac{1}{2}-i\nu +\frac{n}{2}\right)
+
\psi\left(\frac{1}{2}+i\nu -\frac{n}{2}\right)
+
\psi\left(\frac{1}{2}-i\nu -\frac{n}{2}\right)  \nonumber
\right)+2 \psi(1)
\eeqn
\normalsize
while the adjoint BFKL eigenvalue  is given by
\beqn\label{Ehol0}
&&E_{\nu,n}=-\frac{1}{2} \frac{|n|}{\nu^2 +\frac{n^2}{4}}+\psi\left(1+i \nu+\frac{|n|}{2}\right)
+\psi\left(1-i \nu+\frac{|n|}{2}\right)
\\
&& =\frac{1}{2} \left(\psi\left( i\nu +\frac{n}{2}\right)
+
\psi\left(-i\nu +\frac{n}{2}\right)
+
\psi\left( +i\nu -\frac{n}{2}\right)
+
\psi\left( -i\nu -\frac{n}{2}\right)  \nonumber
\right)-2 \psi(1).
\eeqn
This fact suggests that there must be a transformation that effectively takes the BFKL eigenvalue from the singlet to the adjoint representation and probably back, though we believe that  the latter has a big deal   of ambiguity.
At the leading order this transformation is not difficult to find from  (\ref{chi01}) and (\ref{Ehol0})
\beqn\label{subst}
\chi(n,\gamma) \Rightarrow -E_{\nu,n}, \; \psi(M)-\psi(1) \Rightarrow E^{+}_{\nu,n}
, \; \psi(1-\tilde{M})-\psi(1) \Rightarrow E^{-}_{\nu,n},
\eeqn
and at the NLO level it is   less obvious. This transformation   corresponds to  shifting  the absolute value of the conformal spin $|n|$ by minus unity and can be regarded as an analytic continuation of the absolute values of the conformal spin to   negative integers.

As we have already mentioned the singlet case has alternating sums in contrast to the adjoint BFKL, which is  related to the cylindrical topology of the singlet BFKL  versus the planar topology of the adjoint BFKL. In this sense "flattening"  of the singlet BFKL eigenvalue together with the transformation in  (\ref{subst}) is supposed to  transform the singlet BFKL to the adjoint BFKL. By "flattening" we mean converting alternating sums to non-alternating ones  by taking $(-1)^k \to 1$ under the sum over $k$. This transformation we denote by  $\rightarrowtail$.


 For example, consider $\beta'(z)$ in  (\ref{betaprime}) and "flatten" it
\beqn\label{betaprimeflat}
\beta'(z)=-\sum_{k=0}^{\infty} \frac{(-1)^k}{(k+z)^2} \rightarrowtail   - \sum_{k=0}^{\infty} \frac{1}{(k+z)^2} =-\psi'(z).
\eeqn
Next we "flatten" the most complicated function  appearing in the NLO singlet BFKL, namely  $\Phi_2(z)$ in  (\ref{Phi2})
\beqn\label{Phi2flat}
&&\Phi_2(z)=\sum_{k=0}^{\infty} \frac{\beta'(k+1)+ (-1)^k \psi'(k+1)}{k+z}-\sum_{k=0}^{\infty} \frac{ (-1)^k (\psi(k+1)-\psi(1))}{(k+z)^2} \nonumber
\\
&&\rightarrowtail   \sum_{k=0}^{\infty} \frac{-\psi'(k+1)+  \psi'(k+1)}{k+z}-\sum_{k=0}^{\infty} \frac{    \psi(k+1)-\psi(1)  }{(k+z)^2} \nonumber
\\
&&=- \sum_{k=0}^{\infty} \frac{   \psi(k+1)-\psi(1) }{(k+z)^2}
=\frac{\psi''(z)}{2}- \psi'(z)(\psi(z)-\psi(1) ).
\eeqn
Plugging "flattened" $\beta'(z)$ of (\ref{betaprimeflat}) and $\Phi_2(z)$ of (\ref{Phi2flat}) into     (\ref{PhiGamma}) we write
 \beqn\label{PhiGammaflat}
&&\Phi(|n|,\gamma)+\Phi(|n|,1-\gamma) = \Phi_2(M)+\beta'(M)\left[\psi(1)-\psi(1-\tilde{M})\right]  \nonumber
\\
&&
+
\Phi_2(1-\tilde{M})+\beta'(1-\tilde{M})\left[\psi(1)-\psi(M)\right] \nonumber
\\&&
\rightarrowtail \frac{\psi''(M)}{2}+\frac{\psi''(1-\tilde{M})}{2}- \left(\psi(M)-\psi(1-\tilde{M})\right)\left(\psi'(M)-\psi'(1-\tilde{M})\right).
\eeqn
Then we can use this result for the NLO correction to the singlet BFKL in (\ref{delta}) taking into account that "flattened" function should be multiplied by a factor of $1/2$ to compensate "double counting"  caused by "flattening" as can be seen from the following example
\beqn
\beta'(1)=-\sum_{k=0}^{\infty} \frac{(-1)^k}{(k+1)^2} = \frac{1}{2} \left(-\sum_{k=0}^{\infty} \frac{1}{(k+1)^2}\right)=-\frac{\zeta_2}{2}.
\eeqn
The $1/2$ factor  should be introduced only to terms that  originally were written using  alternating series, namely, $\Phi(|n|,\gamma)$ and $\Phi(|n|,1-\gamma)$. The rest of the terms in (\ref{delta}) remain the same and we write
\beqn\label{deltaflat}
&& \delta(n, \gamma)=\psi''(M)+\psi''(1-\tilde{M})+6 \zeta_3-2 \zeta_2 \chi(n,\gamma)-2\Phi(|n|,\gamma)-2\Phi(|n|,1-\gamma) \nonumber
\\
&&
\rightarrowtail\psi''(M)+\psi''(1-\tilde{M})+6 \zeta_3+2 \zeta_2 \left(\psi(M)+\psi(1-\tilde{M})-2 \psi(1)\right) \nonumber
\\
&& -\frac{2}{2} \left(\frac{\psi''(M)}{2}+\frac{\psi''(1-\tilde{M})}{2}- \left(\psi(M)-\psi(1-\tilde{M})\right)\left(\psi'(M)-\psi'(1-\tilde{M})\right)\right) \nonumber
\\
&&
=\frac{1}{2}\left(\psi''(M)+\psi''(1-\tilde{M})\right)+\left(\psi(M)-\psi(1-\tilde{M})\right)\left(\psi'(M)-\psi'(1-\tilde{M})\right) \nonumber
\\
&&+6 \zeta_3+2 \zeta_2 \left(\psi(M)+\psi(1-\tilde{M})-2 \psi(1)\right).
\eeqn
Finally making a substitution we found  in (\ref{subst})
\beqn\label{subst1}
\chi(n,\gamma) \Rightarrow -E_{\nu,n}, \; \psi(M)-\psi(1) \Rightarrow E^{+}_{\nu,n}
, \; \psi(1-\tilde{M})-\psi(1) \Rightarrow E^{-}_{\nu,n},
\eeqn
  together with a similar expression for the  derivatives of the $\psi$-functions~\footnote{The factor $(-1)^m$ in (\ref{der}) reflects the difference in derivatives of $\psi(z)$ denoted by $\psi^{(n)}(z)$ and $D_\nu=-i \partial_\nu$ applied to $\psi(1\pm i \nu +|n|/2)$.}
\beqn\label{der}
\psi^{(m)} (M) \Rightarrow (-1)^m D^m_\nu  E^{+}_{\nu,n}, \;\; \psi^{(m)} (1-\tilde{M}) \Rightarrow   D^m_\nu  E^{-}_{\nu,n}
\eeqn
we obtain
\beqn\label{deltaflatcont}
&& \delta(n, \gamma)
\rightarrowtail \frac{1}{2}\left(\psi''(M)+\psi''(1-\tilde{M})\right)+\left(\psi(M)-\psi(1-\tilde{M})\right)\left(\psi'(M)-\psi'(1-\tilde{M})\right) \nonumber
\\
&&\hspace{2cm}+6 \zeta_3+2 \zeta_2 \left(\psi(M)+\psi(1-\tilde{M})-2 \psi(1)\right) \nonumber
\\
&&
\Rightarrow
-2  \left(
-\frac{1}{4}D^2_{\nu} E_{\nu,n} +\frac{1}{2} \left(E^{+}_{\nu,n}-E^{-}_{\nu,n}\right) D_{\nu} E_{\nu,n}-\zeta_2 E_{\nu,n}-3 \zeta_3
\right).
\eeqn

The last line in (\ref{deltaflatcont}) reproduces the  adjoint NLO eigenvalue  $E^{(1)}_{\nu,n}$ in (\ref{E1Dixon}) provided we replace $V$ by $\tilde{V}=E^{+}_{\nu,n}-E^{-}_{\nu,n}$ in agreement with  the hermitian separability discussed in the previous section. The overall coefficient of $-2$ in (\ref{deltaflatcont}) stems from the difference in the definition of NLO corrections in  (\ref{omegaadjoint}) and (\ref{omegasinglet}) as well as a relative weight of the  color factors for the BFKL eigenvalues in the singlet and adjoint representations.

The fact that in (\ref{deltaflatcont}) we reproduced exactly the modified form of the adjoint NLO BFKL eigenvalue in (\ref{E1tilde})
  supports our proposal that one should consider a modified form of the adjoint NLO BFKL eigenvalue in (\ref{E1tilde}) for checking the  consistency of the adjoint BFKL eigenvalue   with  the general properties of the singlet BFKL, like hermitian separability, bootstrap, $\nu \to 0$ and $\nu \to \infty $ limits.

  We plan to apply the same analysis to the higher order corrections of the BFKL eigenvalue in the adjoint representation and hope that it will help to resolve the problem of non-vanishing  NNLO correction in the limit $\nu \to 0$ for $n=0$~\cite{Dixon4loops}.
   This problem has already been discussed in \cite{CaronHuot2013fea} and shown to be resolved by substituting the factor $1/(\nu^2+n^2/4)$
$1/(\nu^2+n^2/4)$ with $1/(\nu^2+n^2/4-\pi^2 a^2)$ in (\ref{R6}). However, we recall that the factor $1/(\nu^2+n^2/4)$ originated from a product of two impact factors $1/(i\nu+n/2)$ and $1/(-i\nu+n/2)$ standing for two emitted gluons of the same helicity to which the BFKL Green function is attached, and thus  from a point of view   of the Regge factorization the modification $1/(\nu^2+n^2/4-\pi^2 a^2)$ looks quite unnatural. This argument becomes even more transparent if one considers non-MHV amplitude, where a product of two identical terms $1/(i\nu+n/2)$  results in $1/(i\nu+n/2)^2$ that    according to   \cite{CaronHuot2013fea} then need to be modified by $\pi^2 a^2$  to resolve the problem of non-vanishing NNLO eigenvalue in the limit $\nu \to 0$ for $n=0$, while preserving the Regge factorizable structure of the scattering amplitude.
The said above motivates us to search a more natural solution of the non-vanishing adjoint  NNLO BFKL  eigenvalue in the limit $\nu \to 0$ for $n=0$, keeping the original meaning of  $\nu$ and $n$ as labeling parameters of the BFKL eigenfunctions.

\section{Conclusions}
In the present  article, we revisited  the next-to-leading correction to the eigenvalue of  the BFKL equation in the color adjoint representation found by Fadin and Lipatov~\cite{FadinAdj}. This result was confirmed by a number of studies using various calculation techniques and served as a basis for higher order calculations of the adjoint BFKL and the remainder function of the BDS amplitude.  Already at the NNLO level  the adjoint BFKL eigenvalue has an alerting feature of being non-vanishing in the limit of $\nu \to 0$ for $n=0$~\cite{Dixon4loops}, which is in the contradiction with the bootstrap condition. This fact motivated us to go back to the adjoint NLO BFKL eiegnvalue and check its properties against the well studied singlet NLO BFKL eigenvalue. At the leading order both singlet and adjoint BFKL eigenvalues have holomorphic separability, which is broken already at the NLO level. However, it was argued by Kotikov and Lipatov~\cite{KotikovDGLAP} that the singlet NLO BFKL eigenvalue  still possess hermitian separability if written in the form of the Bethe-Salpeter equation.  In the attempt of translating the property of the hermitian separability to the adjoint BFKL eigenvalue we faced a difficulty, which can be resolved by slightly modifying the expression derived in \cite{FadinAdj}. The modification  given by (\ref{Vtilde})   does not affect the bootstrap condition and looks natural if written in terms of holomorphic and antiholomorphic parts of the leading order eigenvalue and their derivatives. The color adjoint BFKL has infrared divergencies in contrast to the well defined singlet BFKL equation and we believe that there should be some ambiguity in  removing those divergencies that can provide a necessary freedom   for modifying the adjoint NLO BFKL eigenvalue as suggested in the present article based on hermitian separability property.
Our suggestion is also supported by the fact that we can  derive a relatively simple and straightforward prescription for passing from the singlet to the adjoint NLO eigenvalue and as a result obtain it exactly in the modified form suggested by the property of the hermitian separability.  This prescription is based on the observation that the major difference between the functions entering the singlet and the adjoint NLO eigenvalue is that the former are related to alternating series whereas the latter to the series of constant sign. It was natural to  assume (see Sec.~\ref{fromto}) that a transformation that  takes the singlet eigenvalue to the adjoint eigenvalue is the one that replaces $(-1)^k$  by unity for the summation over $k$. This assumption happens to be correct at the NLO level and we leave the analysis of  the higher order corrections for future studies. It is worth emphasizing that      the way back from the adjoint to the singlet NLO correction is more ambiguous and does not alow one-to-one correspondence between adjoint and singlet eigenvalues. However, it   can serve a   valuable constraint for future calculations of the higher order corrections to the singlet BFKL equation, which are currently available only at the NLO level for full dependence on the conformal spin $n$ and recently calculated at the NNLO level for $n=0$~\cite{Gromov:2015dfa, veliz}. We have no doubt that the calculation of Fadin and Lipatov in \cite{FadinAdj} are correct, but we argue that the notion of the "adjoint BFKL eigenvalue" should be changed in favor of the modified expression presented in this article and the result in \cite{FadinAdj} only  captures its part relevant for computation of the remainder function of the BDS amplitude.

We argue that using the ambiguity related to redistribution of the NLO corrections between the impact factor and the BFKL kernel related to the arbitrariness of the energy scale, it is possible to bring the NLO eigenvalue to the desired form, while keeping the amplitude and thus the remainder function unchanged.    

In the future we plan to investigate higher order corrections and in particular we hope that this analysis will be helpful in resolving the issue of the non-vanishing adjoint NNLO eigenvalue in the limit of $\nu \to 0$ for $n=0$~\cite{Dixon4loops}, making the result compatible with the bootstrap condition of the BFKL equation.

\section{Acknowledgements}
We are grateful  to  J.~Bartels, V.~S.~Fadin, L.~N.~Lipatov and A.~Sabio Vera for enlightening discussions. A. Prygarin is especially indebted to J.~Bartels for the hospitality during his stay in Hamburg.
The work was supported in part by the SFB   Fellowship.

\end{document}